\documentclass[letter]{aa} 

\usepackage{graphicx}
\usepackage{txfonts}
\usepackage{tabularx, booktabs, array}

\begin{document}

   \title{The massive hot subdwarf binary LAMOST J065816.72+094343.1}

   \author{F. Mattig
          \inst{1}
          \and
          B. N. Barlow \inst{2}
          \and 
          D. Liu \inst{3,4,5}
          \and 
          M. Dorsch \inst{1}
          \and
          S. Geier \inst{1}
          \and
          M. Pritzkuleit \inst{1}
          \and
          H. Dawson \inst{1}
          \and 
          B. Wang \inst{3,4,5}
          \and 
          V. Schaffenroth \inst{6}
          \and           
          T. Kupfer \inst{7,8}
          \and
          C. Derbyshire \inst{2}
          \and 
          S. Barocci-Faul \inst{7}
          }

   \institute{Institut f\"ur Physik und Astronomie, Universit\"at Potsdam, Haus 28, Karl-Liebknecht-Str. 24/25, D-14476
   Potsdam-Golm, Germany
          \and
   Department of Physics and Astronomy, University of North Carolina, 120 E. Cameron Avenue, Chapel Hill, NC 27599, USA
      	 \and
   Yunnan Observatories, Chinese Academy of Sciences, Kunming 650011, PR China
   		 \and
   International Centre of Supernovae (ICESUN), Yunnan Key Laboratory, Kunming 650216, China
          \and
   University of Chinese Academy of Sciences, Beijing 100049, China
          \and
   Thüringer Landessternwarte Tautenburg, Sternwarte 5, D-07778 Tautenburg, Germany
          \and
   Hamburger Sternwarte, University of Hamburg, Gojenbergsweg 112, 21029 Hamburg, Germany
          \and
   Department of Physics and Astronomy, Texas Tech University, Lubbock, TX 79409-1051, USA
             }

   \date{Received  Accepted }

  \abstract
   {Massive short-period binaries involving hot subdwarf stars (sdO/Bs) are rare but very relevant to constraining pathways for binary star evolution. Moreover, some of the most promising candidate progenitor systems leading to Type Ia supernovae (SNe Ia) involve sdO/Bs. LAMOST\,J065816.72+094343.1 has been identified to be such a candidate system.}
   {To explore the nature and evolutionary future of LAMOST\,J065816.72+094343.1, we complemented archival spectroscopic data with additional time series spectra as well as high-resolution spectroscopy of the object. After combining these with photometric data, we determined the orbital parameters of the system and the mass of the companion.}
   {We solved the orbit of the system by analyzing 68 low- and medium-resolution spectra using state-of-the-art mixed local thermodynamic equilibrium (LTE) and non-LTE model atmospheres. Additionally, we gathered nine high-resolution spectra to determine atmospheric parameters and the projected rotational velocity of the sdOB. The inclination angle of the system was constrained assuming tidal synchronization of the sdOB, which was verified via analysis of the ellipsoidal variations in the TESS light curve.}
   {We determine LAMOST\,J065816.72+094343.1 to be a binary consisting of a massive $0.82 \pm 0.17 \, \mathrm{M}_{\odot}$ sdOB component with a $1.30^{+0.31}_{-0.26} \, \mathrm{M}_{\odot}$ unseen companion. Due to the companion's mass being very close to the Chandrasekhar mass limit and high for a white dwarf, it is unclear whether the compact companion is a white dwarf or a neutron star. We find the system to be in a close orbit, with a period of $P=0.31955193 \, \mathrm{d}$ and an inclination angle of $i = 49.6^{+5.2}_{-4.2} \,\mathrm{deg}$. While the exact nature of the companion remains unknown, we determine the system to either lead to a SN Ia or an intermediate mass binary pulsar, potentially after a phase as an intermediate-mass X-ray binary.}
   {}


   \maketitle
\begin{figure*}
    \centering
    \includegraphics[width=.9\textwidth]{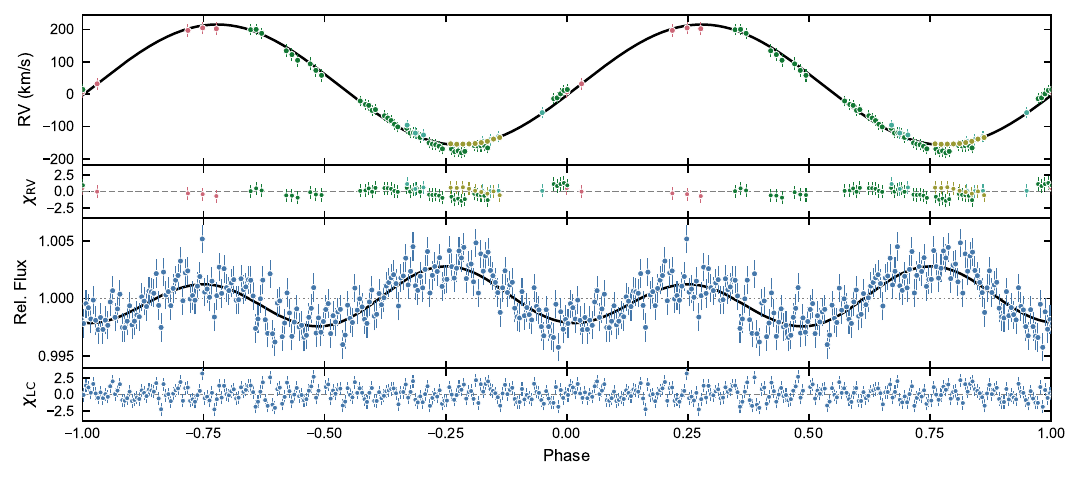}
    \caption{Phased RV curve and phased and binned TESS light curve for J0658. Both curves are phased to the same period, $P$, and zero point, $t_0$ (see Tab.~\ref{tab:Atmospheric Parameters}). For the RV curve, the plots show individual RV measurements from the LAMOST survey (salmon), the SOAR Goodman spectrograph (green; \citealt{2004SPIE.5492..331C}), the ALFOSC spectrograph (light blue), and the UVES spectrograph (ocher). The light curve shows 118287 TESS data points from TESS sectors 33 and 87 that were phased and then binned into 175 bins. Both curves show their respective best-fitting models. For the RV curve, the fit was obtained via the MCMC method, and for the TESS light curve it was obtained using the LCURVE code.}
    \label{fig: curves}
\end{figure*}
\section{Introduction}
Most hot subdwarf stars (sdO/Bs) are evolved core-helium-burning objects that populate the extreme horizontal branch of the Hertzsprung–Russell diagram \citep{2016PASP..128h2001H}. For sdO/Bs to be formed, their progenitors need to evolve up to the red giant phase and subsequently lose all or most of their outer layers on or close to the tip of the red giant branch, a process driven by binary interactions \citep{2002MNRAS.336..449H}. Consequently, about one third of these stars are found in short-period binary systems, with mostly brown dwarf, main-sequence red dwarf, or white dwarf (WD) partners \citep{2022A&A...666A.182S}.

The present-day masses of sdB stars retain information about these binary interactions. Most observed sdB stars have masses clustered around the canonical value for degenerate helium ignition, $M \simeq 0.47\,\mathrm{M}_\odot$, but both lower and higher masses are possible if the progenitor exceeded $\sim 2\,\mathrm{M}_\odot$ and developed a nondegenerate core \citep{2024MNRAS.52711184A}.

Massive short-period sdO/B+WD binaries are particularly interesting because some of them are regarded as potential progenitor systems of type Ia supernovae (SNe Ia; \citealt{2023RAA....23h2001L}). These powerful transient events are triggered when runaway carbon burning ignites in a C/O white dwarf star, and they are very relevant for distance determinations at the largest scales and therefore cosmology. Hot subdwarf binaries qualify for the double-degenerate (DD) SN Ia channel if their total mass is close to or exceeds the Chandrasekhar limit ($M_\mathrm{ch}\approx1.4\,\mathrm{M}_\odot$) and if they are close enough for gravitational radiation to merge them on the scale of a Hubble time. Because an sdO/B turns into a white dwarf on a timescale of only $\sim100\,\mathrm{Myr}$, such systems become WD+WD pairs well before coalescence, given that the post-common envelope period is sufficiently long. To date, three sdO/B+WD systems are known candidates for the super-Chandrasekhar DD merger scenario\footnote{Some sdO/B+WD binaries are SN\,Ia progenitors for other channels as well.} \citep{2007A&A...464..299G, 2021NatAs...5.1052P,2025SCPMA..6869511L}.

Binary-evolution models also predict sdO/Bs with neutron star (NS) or black hole (BH) companions \citep{2018A&A...618A..14W}, and sufficiently short period sdO/B+NS binaries are predicted to be progenitors of low-mass X-ray binaries \citep{2023pbse.book.....T}. Despite theoretical predictions, no sdO/B in a confirmed short-period orbit with an NS or BH companion has been found to date, and existing candidates are either wide binaries or systems with poorly constrained parameters \citep{2011ApJ...743L..11M, 2015A&A...577A..26G, 2023A&A...677A..11G}.

In this paper we report the discovery of the massive short-period sdOB+WD binary \object{LAMOST J065816.72+094343.1} (hereafter J0658). J0658 was first observed by the Large Sky Area Multi-Object Fiber Spectroscopic Telescope (LAMOST) and classified as a sdOB type hot subdwarf by \cite{2018ApJ...868...70L}.
In Section \ref{sec:observations} we present our spectroscopic observations and data reduction. Atmospheric and orbital parameters as well as component masses are derived in Sect.~\ref{sec:analysis}. We discuss the possible evolutionary future of the system in Sect.~\ref{sec:evolution}, and we summarize our conclusions in Sect.~\ref{sec:conclusions}.

\section{Observations}\label{sec:observations}
Five archival LAMOST spectra were available for J0658, with a resolution of $\sim 3.05\, \text{Å}$ and covering a wavelength range of $3700\, \text{Å} - 9000\, \text{Å}$. Spectral follow-up observations for J0658 were then conducted using the Southern Astrophysical Research (SOAR) telescope's Goodman High Throughput Spectrograph (\citeauthor{2004SPIE.5492..331C}, \citeyear{2004SPIE.5492..331C}). In total, 42 spectra with a resolution of $\sim 2.7\, \text{Å}$ and covering a wavelength range of $3600\, \text{Å}-5300\, \text{Å}$ as well as three spectra with a resolution of $\sim 0.58 \, \text{Å}$ and covering $3740\, \text{Å} - 4350\, \text{Å}$ were obtained across five nights between September 2023 and March 2025. All SOAR spectra were taken with an exposure time of 300 seconds and were reduced and barycentrically corrected using the Multi-Instrument Data Input Reducer (MIDIR\footnote{MIDIR is available at https://github.com/Fabmat1/MIDIR}) pipeline. A further nine spectra were taken using the UV-Visual Echelle Spectrograph (UVES) at the ESO Very Large Telescope UT2 (VLT-UT2; \citealt{2000SPIE.4008..534D}) on the night of 1 March, 2025. These spectra were exposed for 300 seconds, yielding a wavelength coverage of $3300\, \text{Å}-4500\, \text{Å}$ in the blue and $5700\, \text{Å}-6650\, \text{Å}$ in the red arm at a resolution of $R=40000$. The UVES spectra were reduced using the EsoReflex pipeline \citep{2013A&A...559A..96F}. Finally, seven additional spectra were taken with the Alhambra Faint Object Spectrograph and Camera (ALFOSC) at the Nordic Optical Telescope (NOT) on 21 April 2025 as well as on the following night. The ALFOSC spectra covered a wavelength range from $3400\, \text{Å}-5350\, \text{Å}$ with a resolution of $2.2\, \text{Å}$ and were also reduced using the MIDIR pipeline. In combination with the archival LAMOST data, these spectra cover a very long baseline of 4801 days.
\begin{figure*}
    \centering
    \includegraphics[width=.95\textwidth]{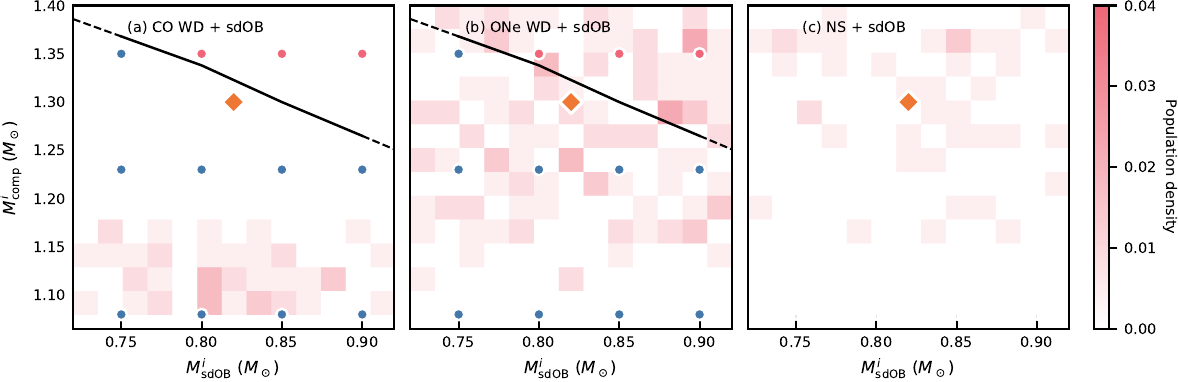}
    \caption{Results of the stellar evolution and BPS simulations. Blue points mark initial configurations producing wide, detached double-WD binaries; red points reach $M_\mathrm{ch}$, triggering an SN Ia or accretion-induced collapse. The background shading shows BPS population densities by companion type, and the orange diamond marks the adopted masses for J0658.}
    \label{fig:future_grid}
\end{figure*}
\section{Analysis}\label{sec:analysis}
\subsection{Atmospheric parameters and radial velocities}\label{sec: Atmospheric Analysis}
We derived the atmospheric parameters by fitting synthetic spectra to the UVES data via $\chi^2$ minimization \citep{2014A&A...565A..63I}. Synthetic spectra were calculated using combined LTE and non-LTE model grids computed with the \textsc{Atlas/Detail/Surface} (ADS) stellar atmosphere and spectral synthesis codes (\citealt{2018A&A...615L...5I} and references therein; Sect. \ref{sec: SpecSol}). Atmospheric parameters from the model fits (Tab.~\ref{tab:Atmospheric Parameters}, Fig. \ref{fig:specfit_UVES}) identify the visible star as a helium-poor sdOB ($T_{\rm eff}=35\,800\pm750\,\mathrm{K}$, $\log g =5.37\pm0.07$, $\log(n_{\rm He})=-1.76\pm0.05$). The spectral fit also yielded a projected rotational velocity of $v\sin i = 37.0\pm1.0 \, \mathrm{km\,s}^{-1}$, which is well constrained due to the metal lines visible in the high-resolution UVES spectra that are modeled in addition to the Balmer and helium lines. A systematic uncertainty of $1.0\, \mathrm{km\,s}^{-1}$ was added in quadrature to the formal uncertainties from the fit.

Radial velocities (RVs) were obtained by fitting the synthetic model template to all the barycentrically corrected spectra. Typical formal fitting uncertainties are $\simeq0.8\,\mathrm{km\,s^{-1}}$ (UVES) and $\simeq5\,\mathrm{km\,s^{-1}}$ (SOAR, ALFOSC, LAMOST). To correct for typical systematic offsets we added $3\,\mathrm{km\,s^{-1}}$ in quadrature to the uncertainties of the UVES RVs and $15\,\mathrm{km\,s^{-1}}$ to the others.

\subsection{Spectral energy distribution}
Fitting the observed spectral energy distribution (SED) using atmospheric priors yields the angular diameter, which we combined with the parallax to derive the stellar radius and mass. The $\chi^2$-minimization method used to construct and fit the SED is described in detail in \cite{2018OAst...27...35H}. For J0658, we adopted the Gaia DR3 parallax and uncertainty, incorporating the zero-point correction of \cite{2021A&A...649A...4L} and the uncertainty adjustment of \cite{2021MNRAS.506.2269E}, and we obtained a precise parallax for the star of $\varpi = 0.79\pm0.05\, \mathrm{mas}$. Interstellar extinction was accounted for using the reddening law of \cite{2019ApJ...886..108F} with $E(44-55)$ as a free parameter and $R(55)=3.02$ (a galactic average). The SED (Fig.~\ref{fig:SED}) does not show an infrared excess. When fitting the SED with the same models as described in Sect. \ref{sec: Atmospheric Analysis}, we found the sdOB to have a large radius, $R_{\mathrm{sd}}=0.309 \pm 0.02 \, \mathrm{R}_{\odot}$, and a high mass, $M_{\mathrm{sd}} = 0.82\pm0.17 \, \mathrm{M}_{\odot}$, compared to other He-poor hot subdwarfs \citep{2025arXiv251102539L}. The color excess measured in the fit ($E(44-55) = 0.1233\pm0.0025\,\mathrm{mag}$) is comparable to the value given by current reddening maps at the position of J0658 ($E(44-55) =0.1042\,\mathrm{mag}$; \citealt{2025Sci...387.1209Z}).

\subsection{Combined spectroscopic and photometric orbital solution}
J0658 was observed in sector 33 and sector 87 of the Transiting Exoplanet Survey Satellite (TESS). We used the PDCSAP flux, which accounts for some crowding present in the TESS aperture (CROWDSAP = 0.64). The resulting light curve with a baseline of 1433 days between the 27-day observations revealed ellipsoidal variations, implying that the hot subdwarf is deformed by the gravitational influence of a compact companion. 

We determined the orbital period by combining information from the light curve and RV data (Tab.~\ref{tab:Radial Velocities}). We calculated the Lomb-Scargle Periodogram \citep{2009A&A...496..577Z} for the TESS light curve and used it as a prior for the period when fitting the RV curve via the Markov chain Monte Carlo (MCMC) approach. For a short-period hot subdwarf binary, tidal forces are expected to efficiently circularize the orbit. We limited the eccentricity to $e<0.04$ in the RV curve and thus adopted $e=0$ for our model, fitting the RVs with $v_{\mathrm{rad}}(t) = K\sin\left({2\pi(t-t_0)}/{P}\right) + \gamma$, where $v_\mathrm{rad}$ is the RV, $K$ is the RV half-amplitude of the sdOB component, $\gamma$ is the systemic velocity, $t_0$ is the zero point of the ephemeris, and $P$ is the orbital period.

This approach yielded a precise period measurement of the system at $P=0.31955193\pm0.00000028\, \mathrm{d}$ ($7.669246\pm0.0000067\, \mathrm{h}$) and a RV half-amplitude of $186.2\pm3.3 \, \mathrm{km\,s}^{-1}$ among the other orbital parameters listed in Tab.~\ref{tab:Atmospheric Parameters}.

\subsection{Companion mass}
We determined the inclination angle and used the binary mass function to derive the mass of the unseen companion. The sdOB in the binary, at its high mass and with its short orbital period, is expected to be tidally synchronized by theory \citep{2024ApJ...975....1M}. We also detected clear light curve variations due to ellipsoidal deformation and Doppler beaming. When comparing the phasing of the light curve with that of the RV curve, we found no phase shift in the tidal bulges, which indicates no deviation from synchronous rotation. We therefore conclude that the rotation of the sdOB in J0658 is synchronized to its orbital motion.

We fit the phased TESS data with the LCURVE code \citep{2010MNRAS.402.1824C} using MCMC sampling. Limb and gravity darkening coefficients used in the fit were sourced from \cite{2020A&A...634A..93C}.
Free parameters were the mass ratio, $q=M_{\mathrm{comp}}/M_{\mathrm{sd}}$; inclination, $i$; velocity scale, $v_s=(1+1/q)K_{\mathrm{sd}}/\sin i$; and scaled sdB radius, $r_{\mathrm{sd}} = R_{\mathrm{sd}}/a = 2\pi R_{\mathrm{sd}}/Pv_s$.  
Priors for all but $i$ were obtained through Monte Carlo sampling of the spectroscopic solution and SED, i.e., independent of the light curve.
The results of the MCMC fit to the light curve can be seen in Fig.~\ref{fig:lc_mcmc}. Due to the very small amplitude of the ellipsoidal modulation and the subsequently low signal-to-noise level of the TESS light curve, the fit only constrains the inclination to the wide range of $i_{\mathrm{LC}} = 50^{+18}_{-12} \,\mathrm{deg}$.

To obtain a constraint on the inclination angle under the assumption of tidal synchronization, we recognized that the rotational velocity of the hot subdwarf in such a system is $v_{\mathrm{rot}} = 2\pi R_{\mathrm{sd}}/P$, where $R_{\mathrm{sd}}$ is the radius of the hot subdwarf and $P$ is the orbital period. Since we had measured $v_{\mathrm{rot}}\sin i$, we could then obtain the inclination angle as $i = \arcsin\left({P\,v_\mathrm{rot}\sin i}/({2\pi R_\mathrm{sd}})\right)$. 
Using this expression with the highly accurate $v_{\mathrm{rot}}\sin i$ obtained with UVES, we obtained a value of $i_{v_{\mathrm{rot}}} = 49.6^{+5.2}_{-4.1} \,\mathrm{deg}$. This inclination angle is consistent with the one obtained from the light curve, but more precise, and we used it for further calculations. Solving the binary mass function numerically with all determined values, we calculated the companion mass to be $M_\mathrm{comp}$ = $1.30^{+0.31}_{-0.26} \,\mathrm{M}_\odot$.

\section{Evolutionary future}\label{sec:evolution}
The unusually massive yet unseen companion of J0658 allows several possible evolutionary pathways, depending on its nature. Its mass is consistent with either an ultramassive CO/ONe WD \citep{2025AN....34640118C} or an NS companion. To explore these possibilities, we modeled the WD companion cases with Eggleton’s stellar evolution code \citep{1971MNRAS.151..351E}, adopting the same assumptions for metallicity, angular-momentum loss, and the Roche-lobe overflow (RLOF) mass transfer as in \citet{2018MNRAS.477..384L}. We computed one reference track with parameters close to the adopted values of the observed system  (Fig. \ref{fig:binary_evolution}) ($M_{\mathrm{i,WD}},\,M_{\mathrm{i,\mathrm{comp}}},\,\log P_{\mathrm{i}})=(1.23 \, \mathrm{M}_\odot,\,0.79 \, \mathrm{M}_\odot,\,-0.4955\,\log(\mathrm{d})$). 

The calculation started with a CO/ONe WD + He–MS binary. After
$\simeq3.3\times10^{7}\,$yr, the He star fills its Roche lobe on the subgiant branch,
initiating RLOF at
$\dot{M}\!\sim\!10^{-7}\,M_\odot\ \mathrm{yr}^{-1}$. The accreted material triggers weak
He shell flashes, steadily increasing the WD mass. After
$2.3\times10^{5}\,$yr, the transfer rate falls below
$6\times10^{-8}\,M_\odot\ \mathrm{yr}^{-1}$, and the flashes become strong, stopping further mass accumulation.  About $3.3\times10^{4}\,$yr later, the donor’s envelope is
exhausted, and it turns into a second WD, leaving a detached double-WD system. At the end of the simulation ($1.9\times10^{5}\,$yr after RLOF), we obtained
$M_{\mathrm{f,WD}}=1.2576\,M_\odot$, $M_{\mathrm{f,\mathrm{comp}}}=0.75\,M_\odot$, and
$\log (P_{\mathrm{f}}/{\rm day})=-0.4594$, values comparable to those of the recently discovered double-WD
SN~Ia progenitor \object{WD\,J181058.67+311940.94} \citep{2025NatAs...9..872M}.
If the accretor is a CO WD, a SN~Ia could follow via the DD channel
in $\sim30\,$Gyr, which is much longer than the Hubble time.

Depending on the specific initial parameter set chosen from the possible parameter space as constrained by the observations, the outcome diverges from the one in this detailed scenario (see Fig. \ref{fig:future_grid}). For combined masses $M_{\mathrm{sys}} \lesssim 2.15\,\mathrm{M}_\odot$, the system does not merge within a Hubble time and forms a wide CO/ONe WD + CO WD binary instead. For CO WD companions at higher masses, the system produces a SN Ia via the DD merger channel at coalescence. For ONe WD companions, the WD undergoes accretion-induced collapse into an NS soon after RLOF, potentially evolving further into an intermediate mass binary pulsar \citep{2018MNRAS.477..384L}. If the companion is already an NS (Fig. \ref{fig:binary_evolution_NS}), stable RLOF during sdOB expansion leads to a transient phase as an intermediate-mass X-ray binary, after which the system becomes an intermediate-mass binary pulsar, as in the ONe case.

To assess the likelihood of the possible companion types we performed Monte Carlo binary population synthesis (BPS) using the Hurley rapid binary evolution code \citep{2002MNRAS.329..897H} with methods similar to those in \cite{2018MNRAS.473.5352L}. We selected J0658-like systems by requiring masses and periods near the observed values ($0.72 < M_\mathrm{sd} < 0.92\,\mathrm{M}_\odot$, $1.08 < M_\mathrm{comp} < 1.4\,\mathrm{M}_\odot$, and $-0.8 < \log P^i < -0.2\,\log(\mathrm{d})$). Results of this simulation can be seen overlaid in Fig. \ref{fig:future_grid}. The BPS results overlap best with the ONe-WD and NS scenarios, and the model explicitly disfavors the SN Ia outcome. For an NS companion, we estimate the maximum current X-ray flux through wind accretion as
$F_{X, \max} \sim10^{-13}\,\mathrm{erg\,cm^{-2}\,s^{-1}}$ (Sect. \ref{sec:Xray_luminosity}), one magnitude larger than the eROSITA limit of $3.2\cdot 10^{-14} \,\mathrm{erg\,cm^{-2}\,s^{-1}}$
at this position \citep{2024A&A...682A..34M}.
The nondetection in eROSITA does not exclude the NS scenario, however, since the calculated maximum flux is idealized and does not factor in detrimental factors such as absorption losses. Overall, the ONe WD or NS companion scenarios can currently be seen as the most likely based on the BPS results.
\section{Conclusions}\label{sec:conclusions}
We have identified and characterized J0658 as a new, highly massive short-period hot subdwarf binary. Atmospheric analysis of the UVES spectra showed the visible component to be a helium-poor sdOB star with $T_{\rm eff}=35\,800\pm750\,\mathrm{K}$, $\log g = 5.37 \pm 0.07$, and a projected rotational velocity of $v\sin i = 37.0 \pm 1.0\,\mathrm{km\,s^{-1}}$. The likely tidally locked nature of the system allowed for a precise measurement of the inclination angle and a subsequent determination of the companion mass. With a primary mass of $M_{\mathrm{sd}} = 0.82\pm 0.17 \, \mathrm{M}_\odot$ and a companion mass of $M_{\mathrm{comp}}=1.30^{+0.31}_{-0.26} \, \mathrm{M}_\odot$, it is the most massive of the now four known potential DD merger SNe Ia progenitor systems.

The companion’s mass range overlaps with those of the most massive CO and ONe white dwarfs and extends into the low-mass neutron-star regime. Notably, the current data do not allow for an unambiguous identification among these possibilities. If the companion is a CO WD, binary-evolution calculations predict stable He accretion and a double-WD merger within $\sim30\,$Gyr, leading to a SN\,Ia or the formation of a detached double WD.  
An ONe WD companion would instead lead to accretion-induced collapse and the formation of an intermediate-mass binary pulsar, which is the same outcome as expected for an NS companion, though the system would first go through a transient phase as a intermediate-mass X-ray binary in this case.

\bibliographystyle{aa}
\bibliography{refs}
\begin{appendix}
\section{Acknowledgements}
\begin{acknowledgements}
    F. M. , H. D., and M. P. are supported by the Deutsche Forschungsgemeinschaft through grants GE2506/17-1, GE2506/9-2 and GE2506/18-1, respectively.
    B.B. and C.D. are supported by faculty research start-up funds kindly provided by the UNC Department of Physics \& Astronomy and the UNC College of Arts \& Sciences.
    This research was supported by Deutsche Forschungsgemeinschaft  (DFG, German Research Foundation) under Germany’s Excellence Strategy - EXC 2121 "Quantum Universe" – 390833306. Co-funded by the European Union (ERC, CompactBINARIES, 101078773). Views and opinions expressed are however those of the author(s) only and do not necessarily reflect those of the European Union or the European Research Council. Neither the European Union nor the granting authority can be held responsible for them.
    This study is supported by the National Natural Science Foundation of China (Nos 12288102, 12225304, 12090040/12090043, 12273105), the National Key R\&D Program of China (No. 2021YFA1600404), the Youth Innovation Promotion Association of the Chinese Academy of Sciences (No. 2021058), the Yunnan Revitalization Talent Support Program (Yunling Scholar Project and Young Talent Project), the Yunnan Fundamental Research Project (No 202201BC070003, 202401AV070006, and 202201AW070011), and the International Centre of Supernovae, Yunnan Key Laboratory (No. 202302AN360001).
    Based on observations obtained at the Southern Astrophysical Research (SOAR) telescope, which is a joint project of the Minist\'{e}rio da Ci\^{e}ncia, Tecnologia e Inova\c{c}\~{o}es (MCTI/LNA) do Brasil, the US National Science Foundation’s NOIRLab, the University of North Carolina at Chapel Hill (UNC), and Michigan State University (MSU). 
    The data presented here were obtained in part with ALFOSC, which is provided by the Instituto de Astrofisica de Andalucia (IAA) under a joint agreement with the University of Copenhagen and NOT. Based on observations collected at the European Southern Observatory under ESO programme 114.28KP.001.
    This paper includes data collected with the TESS mission, obtained from the MAST data archive at the Space Telescope Science Institute (STScI). Funding for the TESS mission is provided by the NASA Explorer Program. STScI is operated by the Association of Universities for Research in Astronomy, Inc., under NASA contract NAS 5–26555. 
    The Pan-STARRS1 Surveys (PS1) and the PS1 public science archive have been made possible through contributions by the Institute for Astronomy, the University of Hawaii, the Pan-STARRS Project Office, the Max-Planck Society and its participating institutes, the Max Planck Institute for Astronomy, Heidelberg and the Max Planck Institute for Extraterrestrial Physics, Garching, The Johns Hopkins University, Durham University, the University of Edinburgh, the Queen's University Belfast, the Harvard-Smithsonian Center for Astrophysics, the Las Cumbres Observatory Global Telescope Network Incorporated, the National Central University of Taiwan, the Space Telescope Science Institute, the National Aeronautics and Space Administration under Grant No. NNX08AR22G issued through the Planetary Science Division of the NASA Science Mission Directorate, the National Science Foundation Grant No. AST–1238877, the University of Maryland, Eotvos Lorand University (ELTE), the Los Alamos National Laboratory, and the Gordon and Betty Moore Foundation. 
    This publication makes use of data products from the Two Micron All Sky Survey, which is a joint project of the University of Massachusetts and the Infrared Processing and Analysis Center/California Institute of Technology, funded by the National Aeronautics and Space Administration and the National Science Foundation.
    The national facility capability for SkyMapper has been funded through ARC LIEF grant LE130100104 from the Australian Research Council, awarded to the University of Sydney, the Australian National University, Swinburne University of Technology, the University of Queensland, the University of Western Australia, the University of Melbourne, Curtin University of Technology, Monash University and the Australian Astronomical Observatory. SkyMapper is owned and operated by The Australian National University's Research School of Astronomy and Astrophysics. The survey data were processed and provided by the SkyMapper Team at ANU. The SkyMapper node of the All-Sky Virtual Observatory (ASVO) is hosted at the National Computational Infrastructure (NCI). Development and support of the SkyMapper node of the ASVO has been funded in part by Astronomy Australia Limited (AAL) and the Australian Government through the Commonwealth's Education Investment Fund (EIF) and National Collaborative Research Infrastructure Strategy (NCRIS), particularly the National eResearch Collaboration Tools and Resources (NeCTAR) and the Australian National Data Service Projects (ANDS).
    This work presents results from the European Space Agency (ESA) space mission Gaia. Gaia data are being processed by the Gaia Data Processing and Analysis Consortium (DPAC). Funding for the DPAC is provided by national institutions, in particular the institutions participating in the Gaia MultiLateral Agreement (MLA). The Gaia mission website is \url{https://www.cosmos.esa.int/gaia}. The Gaia archive website is \url{https://archives.esac.esa.int/gaia}.\\\\
    \hspace{0pt}
\end{acknowledgements}
\section{Supplementary tables}
\begin{table}[!h]
    \caption{Parameters of LAMOST\,J065816.72+094343.1.}
    \label{tab:Atmospheric Parameters}
    \centering
    \small
    \begin{tabularx}{0.85\columnwidth}{@{}X r@{}}
        \toprule\toprule
        \multicolumn{2}{c}{Gaia DR3 astrometry} \\
        \midrule
        
        RA (J2000)                        & 06:58:16.72\\
        Dec (J2000)                       & +09:43:43.12\\
        $G$ (mag)                         & 13.57\\
        $\varpi$ (mas)                    & $0.79\pm0.05$\\
        RUWE                              & 1.085\\
        \midrule
        \multicolumn{2}{c}{Spectroscopic fit} \\
        \midrule
        
        $T_{\rm eff}$ (K)                 & $35\,800\pm750$\\
        $\log g$ (dex)                    & $5.37\pm0.07$\\
        $\log(n_{\rm He})$                & $-1.76\pm0.05$\\
        $\log(n_{\rm N})$                 & $-4.1\pm0.1$\\
        $\log(n_{\rm Si})$                & $-5.0\pm0.1$\\
        $v_{\rm rot}\sin i$ (km\,s$^{-1}$)& $37.0\pm1.0$\\
        \midrule
        \multicolumn{2}{c}{SED fit} \\
        \midrule
        
        $R_{\rm sd} = \Theta/2\varpi$ ($R_\odot$)          & $0.309\pm0.020$\\
        $M_{\rm sd} = gR^2/G$ ($M_\odot$)          & $0.82\pm0.17$\\
        $L_{\rm sd} = 4\pi R^2\sigma_{\mathrm{SB}}T_{\mathrm{eff}}^4$ ($L_\odot$)          & $139^{+22}_{-19}$\\
        $E(44-55) (\mathrm{mag})$         & $0.1233\pm0.0025$\\
        \midrule
        \multicolumn{2}{c}{RV fit (median values)} \\
        \midrule
        
        $P$ (d)                           & $0.319\,551\,93\pm0.000\,000\,28$\\
        $K$ (km\,s$^{-1}$)                & $186.2\pm3.3$\\
        $f$ ($M_\odot$)                   & $0.2137\pm0.011$\\
        $\gamma$ (km\,s$^{-1}$)           & $29.4\pm2.9$\\
        $t_0$ (BJD$_{\rm UTC}$)           & $2\,455\,939.1792\pm0.0038$\\\midrule
        \multicolumn{2}{c}{Light curve fit (median values)} \\
        \midrule
        
        $i_{\rm LC}$ (deg)                & $49^{+19}_{-15}$\\
        \addlinespace[0.5ex] 
        $q = M_{\rm comp}/M_{\rm sd}$     & $1.57^{+0.92}_{-0.50}$\\
        \addlinespace[0.5ex] 
        $r_{\rm sd} = R_{\rm sd}/a$       & $0.118^{+0.011}_{-0.012}$\\
        \addlinespace[0.5ex] 
        $v_{\rm s}$ (km\,s$^{-1}$)       & $427^{+65}_{-57}$\\
        \midrule
        \multicolumn{2}{c}{Derived} \\
        \midrule
        
        $i_{v_{\rm rot}}$ (deg)           & $49.6^{+5.2}_{-4.1}$\\
        \addlinespace[0.5ex] 
        $M_{\rm comp}$ ($M_\odot$)        & $1.30^{+0.31}_{-0.26}$\\
        \bottomrule
    \end{tabularx}
\end{table}
\begin{table}[!h]
\caption{Measured RVs. The 1$\sigma$ uncertainties are indicated under $\sigma_{\rm RV}$.}
\label{tab:Radial Velocities}
\centering
\small
\begin{tabular}{cccc}
\toprule\toprule
RV (km/s) & $\sigma_{\rm RV}$ (km/s) & BJD$_{\rm UTC}$ & Instrument \\
\midrule
6.79 & 19.13 & 2455939.18855 & LAMOST \\
32.05 & 18.94 & 2455939.19827 & LAMOST \\
197.15 & 19.40 & 2457385.23076 & LAMOST \\
204.85 & 18.81 & 2457385.24048 & LAMOST \\
202.30 & 19.41 & 2457385.24951 & LAMOST \\
-20.97 & 16.75 & 2460330.65476 & SOAR \\
-31.81 & 17.09 & 2460330.65820 & SOAR \\
-34.54 & 16.24 & 2460330.66040 & SOAR \\
-50.05 & 16.90 & 2460330.66260 & SOAR \\
-47.58 & 16.43 & 2460330.66480 & SOAR \\
-66.75 & 16.65 & 2460330.67074 & SOAR \\
-73.58 & 16.61 & 2460330.67294 & SOAR \\
-81.51 & 16.69 & 2460330.67514 & SOAR \\
-92.95 & 16.20 & 2460330.67734 & SOAR \\
-101.08 & 16.64 & 2460330.67954 & SOAR \\
-105.38 & 16.87 & 2460330.68545 & SOAR \\
-120.04 & 16.64 & 2460330.68765 & SOAR \\
-119.13 & 17.22 & 2460330.68985 & SOAR \\
-120.56 & 16.81 & 2460330.69205 & SOAR \\
-133.60 & 17.12 & 2460330.69425 & SOAR \\
-150.35 & 16.90 & 2460330.70020 & SOAR \\
-154.17 & 17.08 & 2460330.70240 & SOAR \\
-156.24 & 16.99 & 2460330.70460 & SOAR \\
-160.88 & 17.29 & 2460330.70680 & SOAR \\
-168.87 & 17.47 & 2460330.70900 & SOAR \\
-169.17 & 17.26 & 2460330.71496 & SOAR \\
-177.86 & 17.47 & 2460330.71716 & SOAR \\
-175.23 & 17.32 & 2460330.71935 & SOAR \\
-181.15 & 17.54 & 2460330.72155 & SOAR \\
-176.76 & 17.86 & 2460330.72375 & SOAR \\
-158.91 & 17.30 & 2460330.73003 & SOAR \\
-159.96 & 17.38 & 2460330.73223 & SOAR \\
-159.54 & 18.12 & 2460330.73443 & SOAR \\
-158.04 & 17.51 & 2460330.73662 & SOAR \\
-166.06 & 16.84 & 2460330.73882 & SOAR \\
-13.89 & 17.36 & 2460705.61684 & SOAR \\
-11.38 & 17.27 & 2460705.61909 & SOAR \\
1.20 & 16.82 & 2460705.62133 & SOAR \\
11.81 & 16.78 & 2460705.62357 & SOAR \\
14.41 & 17.34 & 2460705.62581 & SOAR \\
-153.34 & 3.70 & 2460736.54539 & UVES \\
-154.58 & 3.68 & 2460736.54942 & UVES \\
-154.12 & 3.68 & 2460736.55346 & UVES \\
-153.70 & 3.64 & 2460736.55751 & UVES \\
-152.37 & 3.62 & 2460736.56155 & UVES \\
-149.49 & 3.64 & 2460736.56559 & UVES \\
-145.70 & 3.67 & 2460736.56964 & UVES \\
-138.63 & 3.68 & 2460736.57368 & UVES \\
-133.91 & 3.71 & 2460736.57772 & UVES \\
199.25 & 16.41 & 2460740.56763 & SOAR \\
200.22 & 16.44 & 2460740.57127 & SOAR \\
188.18 & 16.63 & 2460740.57490 & SOAR \\
134.20 & 19.79 & 2460740.59114 & SOAR \\
122.42 & 19.96 & 2460740.59488 & SOAR \\
104.89 & 19.74 & 2460740.59862 & SOAR \\
93.64 & 20.32 & 2460740.60694 & SOAR \\
73.62 & 20.15 & 2460740.61067 & SOAR \\
58.35 & 20.96 & 2460740.61441 & SOAR \\
-95.59 & 17.35 & 2460786.36679 & ALFOSC \\
-120.09 & 17.50 & 2460786.37220 & ALFOSC \\
-126.01 & 17.25 & 2460786.37758 & ALFOSC \\
-145.71 & 18.08 & 2460786.41640 & ALFOSC \\
-140.53 & 18.64 & 2460786.42176 & ALFOSC \\
-130.91 & 17.78 & 2460786.42712 & ALFOSC \\
-56.97 & 17.47 & 2460787.41457 & ALFOSC \\
\bottomrule
\end{tabular}
\end{table}
\hspace{0pt}
\section{Supplementary calculations}
\subsection{Estimated X-ray luminosity}\label{sec:Xray_luminosity}
In order to constrain the nature of the unseen companion, we estimate the current X-ray luminosity of the binary for the scenario of a WD and NS companion, and compare it with the point-source detection limit of eROSITA. We use the best-fit mass-loss rate function determined by \cite{2016A&A...593A.101K} to estimate the wind mass-loss rate $\dot{M}_w$ of the hot subdwarf star as 
\begin{equation}
    \dot{M}_w = \left(3.59\pm1.72\right) \cdot 10^{-12} \, \mathrm{M}_{\odot}\mathrm{yr}^{-1}.
\end{equation}
Using a terminal wind velocity of the star estimated from the model values in \cite{2016A&A...593A.101K} as $v_{w, \infty} \approx 1420 \, \mathrm{km\, s}^{-1}$ together with the Bondi–Hoyle–Lyttleton accretion rate and the best-fit determined system parameters, we obtained
\begin{align}
    \dot{M}_{\mathrm{acc}} &= \alpha {\frac {4\pi(GM_{\mathrm{comp}})^{2}\rho }{v_{w, \infty}^{3}}} \approx \frac { G^2M_{\mathrm{comp}}^{2} \dot{M}_w}{a^2v_{w, \infty}^{4}}\\
    &= (8.58 \pm 6.16) \cdot 10^{-15}\, \mathrm{M}_{\odot}\mathrm{yr}^{-1}.
\end{align}
The corresponding X-ray luminosity can be approximated using the accretion rate-luminosity relationship (\citealt{2002apa..book.....F}; Eq. 1.5) and approximating $R_\mathrm{NS} \sim 12\,\mathrm{km}$ and $R_\mathrm{WD}~\sim~3000 \, \mathrm{km} $ based on neutron star equation of state constraints from \cite{2016ARA&A..54..401O} and the well-known empirical mass-radius relationship for white dwarfs by \cite{1972ApJ...175..417N}:
\begin{gather}
    L_\mathrm{X,\,NS} \approx GM_\mathrm{comp}\dot{M}_{\mathrm{acc}}/R_{\mathrm{comp}} = (7.7\pm 5.8)\cdot10^{31} \, \mathrm{erg}/\mathrm{s} \\
    L_\mathrm{X,\,WD} \approx (3.1\pm 2.3)\cdot10^{29} \, \mathrm{erg}/\mathrm{s}, 
\end{gather}
which, at a distance of $1266\pm80\, \mathrm{pc}$ (as calculated from the adopted parallax) corresponds to an ideal flux of
\begin{gather}
F_\mathrm{X,\,NS} = \frac{L}{4 \pi \, d^2} = (4.0 \pm 3.1) \times 10^{-13} \ \mathrm{erg\,cm^{-2}\,s^{-1}}\\
F_\mathrm{X,\,WD} = (1.60 \pm 1.22) \times 10^{-15} \ \mathrm{erg\,cm^{-2}\,s^{-1}}
\end{gather}
for J0658. This calculation does not include potentially detrimental effects such as absorption through neutral hydrogen, which means that the true flux of the system is likely lower than this value.
\subsection{Reliability of the spectroscopic solution} \label{sec: SpecSol}
The reliability of the atmospheric parameters we determine through model fitting is central to our analysis of J0658. We thus give some additional details on our method here and compare our results to the literature.\\
The synthetic spectra were calculated using a state-of-the-art hybrid LTE and non-LTE approach using the ADS code and are described in detail by \cite{moeller2021elemental}. In addition to hydrogen and helium lines, 24 additional elements are modeled in these synthetic spectra, though only \ion{N}{ii}, \ion{Si}{iii} and \ion{Si}{iv} metal lines are actually present in the UVES spectra with equivalent widths of $>10\,\mathrm{m}\text{Å}$. The atmospheric parameters were determined via $\chi^2$ minimization by fitting these synthetic spectra simultaneously to all individual UVES data, while keeping a common set of atmospheric parameters.\\
\cite{2021ApJS..256...28L} previously classified J0658 as an sdOB-type hot subdwarf, finding a divergent surface gravity ($\log g\,_{\rm Luo+} = 5.23$). Their analysis is based on a single co-added low-resolution spectrum from LAMOST DR7 with a combined exposure time of 45 minutes. Since the exposure time point of this spectrum coincides with the RV curve's turnaround point, it is affected by significant smearing due to an RV change of $\sim36\,\mathrm{kms}^{-1}$. When fitting the individual LAMOST exposures making up the co-added spectrum \cite{2021ApJS..256...28L} used with our models and method, we get atmospheric parameters that are closely aligned with our UVES-based solution ($T_{\rm eff}=37000\pm1000\,\mathrm{K}$, $\log g =5.32\pm0.08$, $\log(n_{\rm He})=-1.76\pm0.07$). The discrepancy in derived parameters can therefore be attributed to the differing models, fitting methods and data quality between the two works.
\begin{onecolumn}
\section{Supplementary figures}
\begin{figure}[!h]
\centering
\includegraphics[width=\textwidth]{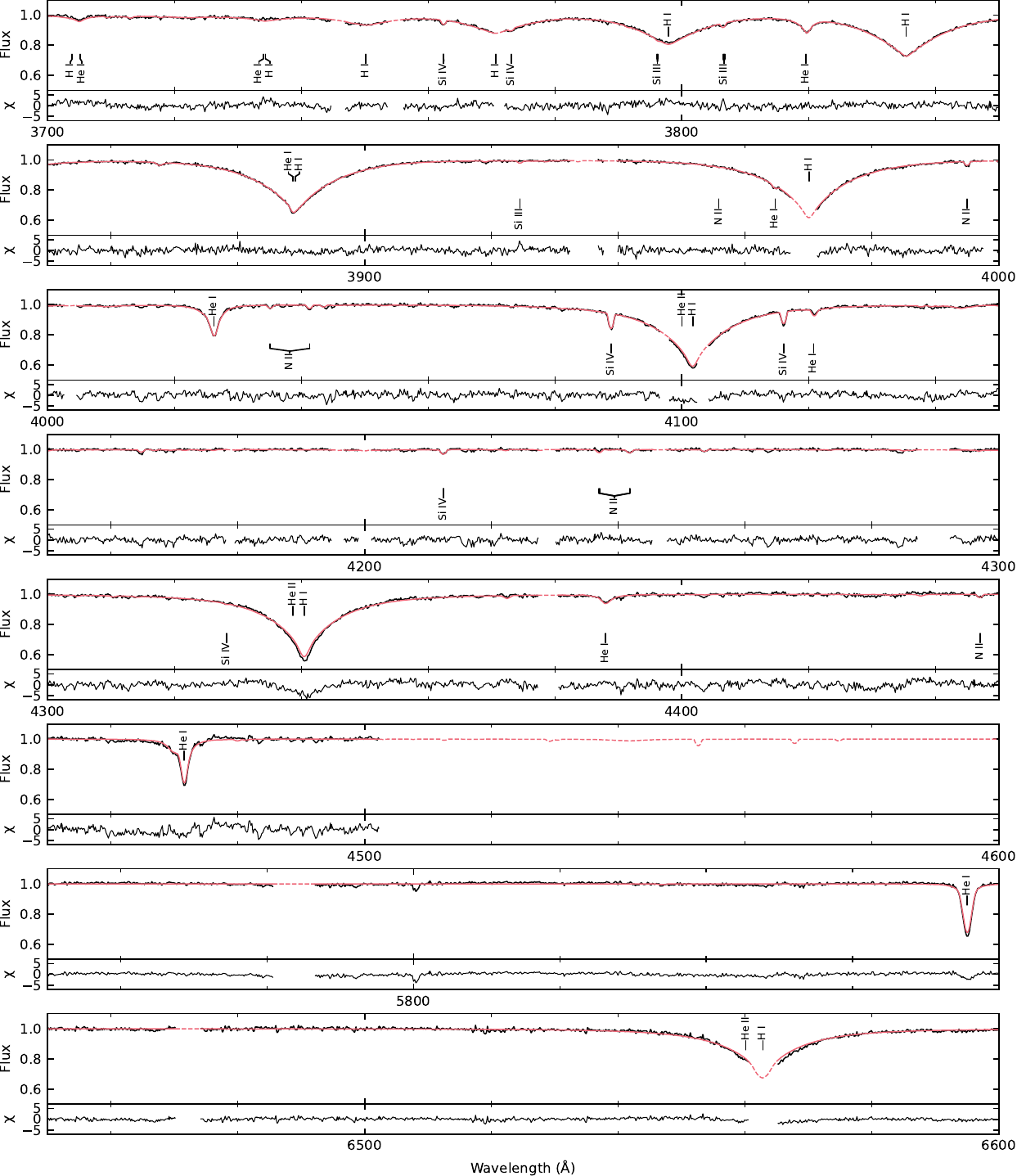}
\caption{Stacked and re-binned (50000 bins) UVES spectra for J0658. The black line represents the spectral data, the red line the best-fitting model spectrum. The gap between the end of the UVES blue arm and red arm ($4500\text{Å} - 5700\text{Å}$) and sections of the spectrum without any absorption lines are not shown.}
\label{fig:specfit_UVES}
\end{figure}
\end{onecolumn}
\twocolumn
\begin{figure*}[!t]
\centering
\begin{minipage}[t]{0.48\textwidth}
   \centering
   \includegraphics[width=\textwidth]{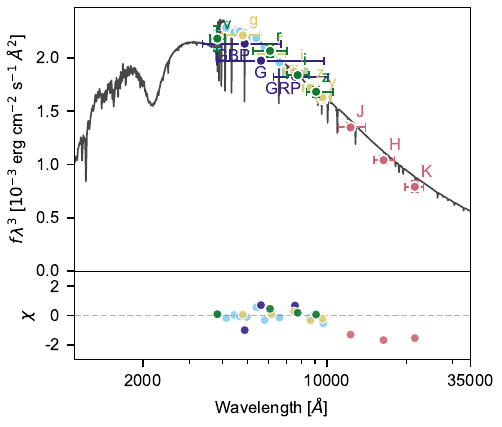}
   \caption{Spectral energy distribution of J0658. Valid fluxes from all available photometric surveys are shown. The best-fitting synthetic spectral model is overlaid as a gray line. Horizontal error bars represent the filter width at tenth maximum. (Data publications: Pan-STARRS1 (yellow): \citealt{2016arXiv161205560C}; 2MASS (salmon): \citealt{2003tmc..book.....C}; SkyMapper (green): \citealt{2019PASA...36...33O}; and Gaia G/BP/RP (blue) and Gaia XP (light blue): \citealt{2022yCat.1355....0G}).}
   \label{fig:SED}
\end{minipage}
\hfill
\begin{minipage}[t]{0.48\textwidth}
    \centering
    \includegraphics[width=\textwidth]{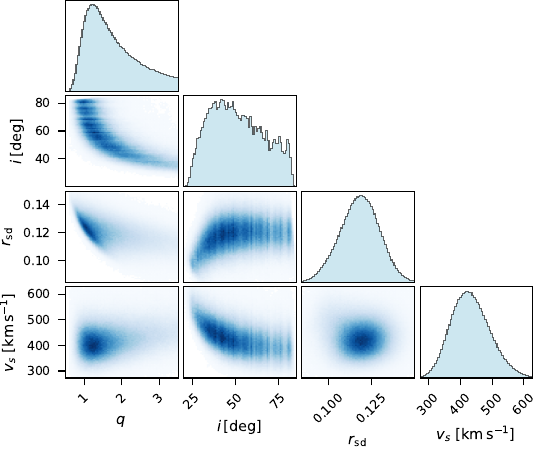}
    \caption{Posterior distributions of the free parameters in the light curve fit, obtained via MCMC sampling with 10\,000\,000 samples of the TESS light curve of J0658.}
    \label{fig:lc_mcmc}
\end{minipage}
\end{figure*}
\begin{figure*}[!h]
\centering
\includegraphics[width=.9\textwidth]{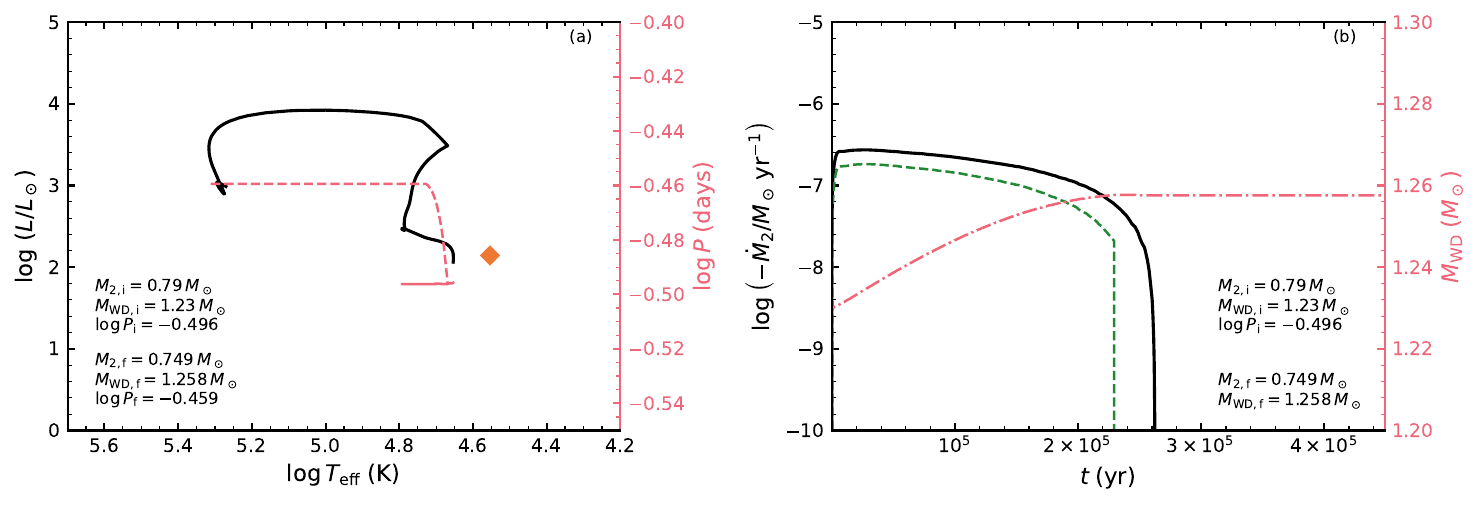}
\caption{Binary evolution model for J0658 assuming a CO–white-dwarf companion. 
Left: Hertzsprung–Russell track of the helium star (solid line) and simultaneous change of the orbital period (dash-dotted line) as well as the current position of J0658 (orange diamond). 
Right: Mass-transfer history after the onset of Roche lobe overflow. Solid, dashed and dash-dotted curves denote the donor’s mass-loss rate, the WD accretion rate, and the growing WD mass, respectively. Insets list the adopted initial parameters and the final values reached at the end of the computation.}
\label{fig:binary_evolution}
\end{figure*}
\begin{figure*}[!h]
\centering
\includegraphics[width=.9\textwidth]{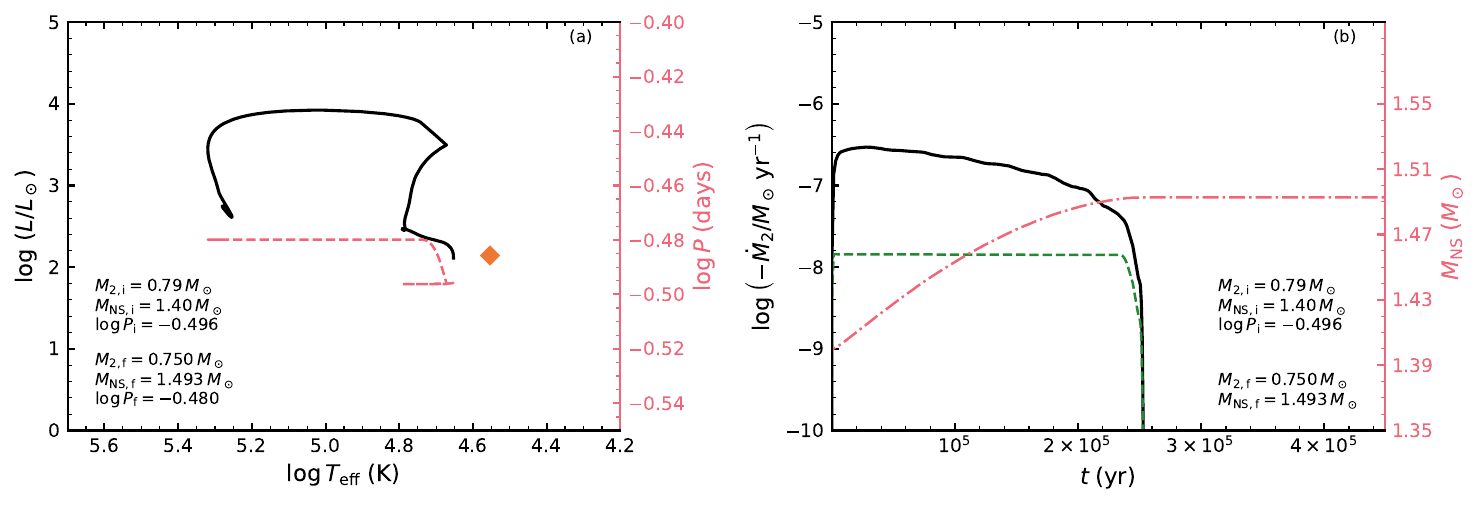}
\caption{Binary evolution model for J0658 assuming an NS companion. 
Contents are analogous to those in Fig. \ref{fig:binary_evolution}.}
\label{fig:binary_evolution_NS}
\end{figure*}

\end{appendix}

\end{document}